\documentstyle[sprocl,axodraw]{article}

\input{epsf}
\def\Journal#1#2#3#4{{#1} {\bf #2}, #3 (#4)}

\def\PLB{{\em Phys. Lett.}  B}

\def\PRD{{\em Phys. Rev.} D}
\def\ZPC{{\em Z. Phys.} C}

\def\be{\begin{equation}}
\def\ee{\end{equation}}
\def\bea{\begin{eqnarray}}
\def\eea{\end{eqnarray}}

\begin{document}

\title{MEASURING THE HIGGS YUKAWA COUPLINGS AT A NEXT LINEAR COLLIDER}

\author{S.~DAWSON }

\address{Physics Department, Brookhaven National Laboratory, Upton, NY 11973}

\author{L.~REINA }

\address{Physics Department, Florida State University, Tallahassee, FL 32306}


\maketitle\abstracts{ We investigate the inclusive production of a
Higgs boson with a pair of heavy quarks ($t\bar t$ or $b\bar b$), in
$e^+e^-$ collisions at high energies, $\sqrt{s}\!=\!500$ GeV and
$\sqrt{s}\!=\!1$ TeV. We consider both the Standard Model and the
supersymmetric case. In both cases $O(\alpha_s)$ QCD corrections are
included. The associated production of a Higgs boson with a $t\bar t$
pair is extremely sensitive to the top-Higgs Yukawa coupling and may
allow the precision measurement of this coupling. In some regions of
the supersymmetric parameter space the associated production of a
Higgs boson with a $b\bar b$ pair receives large resonant
contributions and can have a significant rate.}
  
\section{Introduction}
\label{sec:intro}
The origin and hierarchy of particle masses remains one of the most
tantalizing problems in contemporary particle physics.  The untangling
of the electroweak symmetry breaking is among the very important and
challenging goals of the present and next generation of colliders.  In
this context the search for the Higgs boson and the study of its
couplings to gauge bosons and fermions are crucial. 

Recent bounds from LEP2 have pushed the Higgs mass up to:
$M_\phi\!>\!95.2$ GeV (Standard Model, $\phi\!=\!H_{SM}$)~\cite{read}
and $M_\phi\!>\!81$ GeV (minimal Supersymmetry, $\phi\!=\!h^0,A^0$,
$\tan\beta\!>\!0.5$)~\cite{gay}. At the same time, there are strong
theoretical reasons to believe that, if our understanding of the
electroweak symmetry breaking is correct, a light mass Higgs boson
ought to exist. This is definitely possible in the Standard Model and
mandatory in the minimal version of Supersymmetry, where the upper
bound on the mass of the light Higgs scalar is around 130~GeV. We
therefore focus on the possibility that a so called \emph{intermediate
mass Higgs} exists, with a mass in the range between 100-130 GeV.

If such Higgs boson exists, chances are that it will be discovered
maybe at the present (LEP2) and for sure at the future generation of
colliders (Tevatron RunII or LHC). A high energy Next Linear Collinear
(NLC) would then be the ideal environment for precision studies.  In
this context the $e^+e^-\rightarrow t\bar t\phi$ and
$e^+e^-\rightarrow b\bar b\phi$ production modes can offer the unique
possibility of a direct measurement of the top ($g_{t\bar t\phi}$) and
bottom ($g_{b\bar b\phi}$) Yukawa couplings, both at
$\sqrt{s}\!=\!500$ GeV and at $\sqrt{s}\!=\!1$ TeV.  The production
rates are small (of the order of a few femtobarns), but the signatures
can be quite spectacular, in particular for the $t\bar t\phi$ case.

We have studied the inclusive associated production of both $t\bar
t\phi$ and $b\bar b\phi$ at high energy $e^+e^-$ colliders
($\sqrt{s}\!=\!500$ GeV and $\sqrt{s}\!=\!1$ TeV), including
$O(\alpha_s)$ QCD corrections~\cite{sl_qcd,sl_susy}. QCD corrections
to $e^+e^-\rightarrow t\bar tH_{SM}$ have also been computed in
reference~\cite{ditt}. Both in the Standard Model and in
Supersymmetry, the $e^+e^-\rightarrow t\bar t\phi$ production mode
turns out to be extremely sensitive to the top Yukawa coupling
$g_{t\bar t\phi}$, allowing a very precise measurement of the coupling
itself. We discuss the $t\bar t\phi$ case in Section \ref{sec:ttbar}.
On the other hand, the $e^+e^-\rightarrow b\bar b\phi$ production mode
has a negligible rate in the Standard Model, but gets enhanced by
resonance effects in some regions of the supersymmetric parameter
space. Therefore it can provide evidence of non standard physics. We
discuss the $b\bar b\phi$ case in Section \ref{sec:bbbar}. Section
\ref{sec:concl} contains our conclusions.

\section{Associated production of a Higgs boson with a $t\bar t$ pair}
\label{sec:ttbar}
\subsection{Standard Model}
\label{subsec:ttbar_sm}

\begin{figure}[hbtp]
\centering
\epsfysize=1.5in
\leavevmode\epsffile{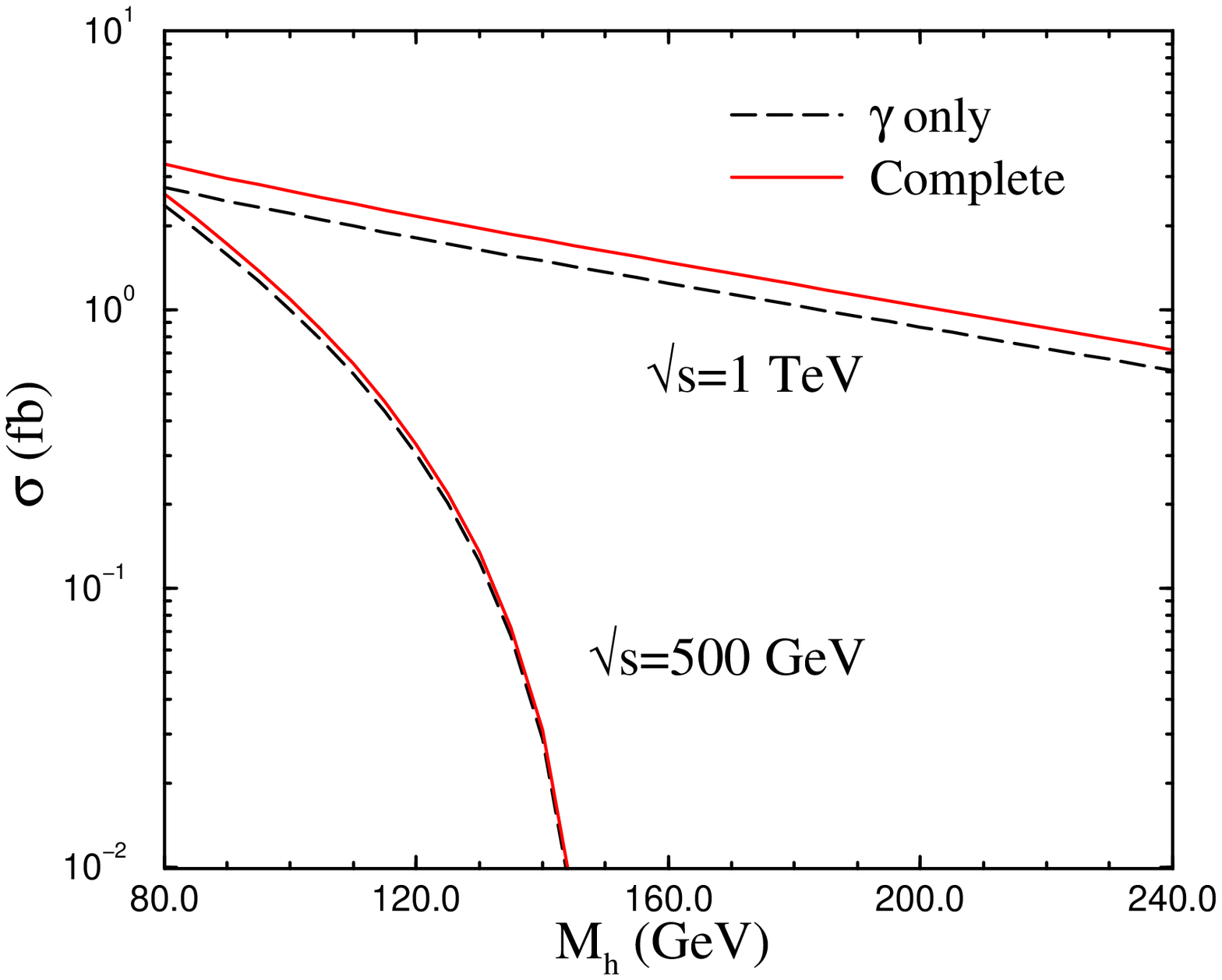}
\hspace{2.truecm}
\epsfysize=1.5in
\leavevmode\epsffile{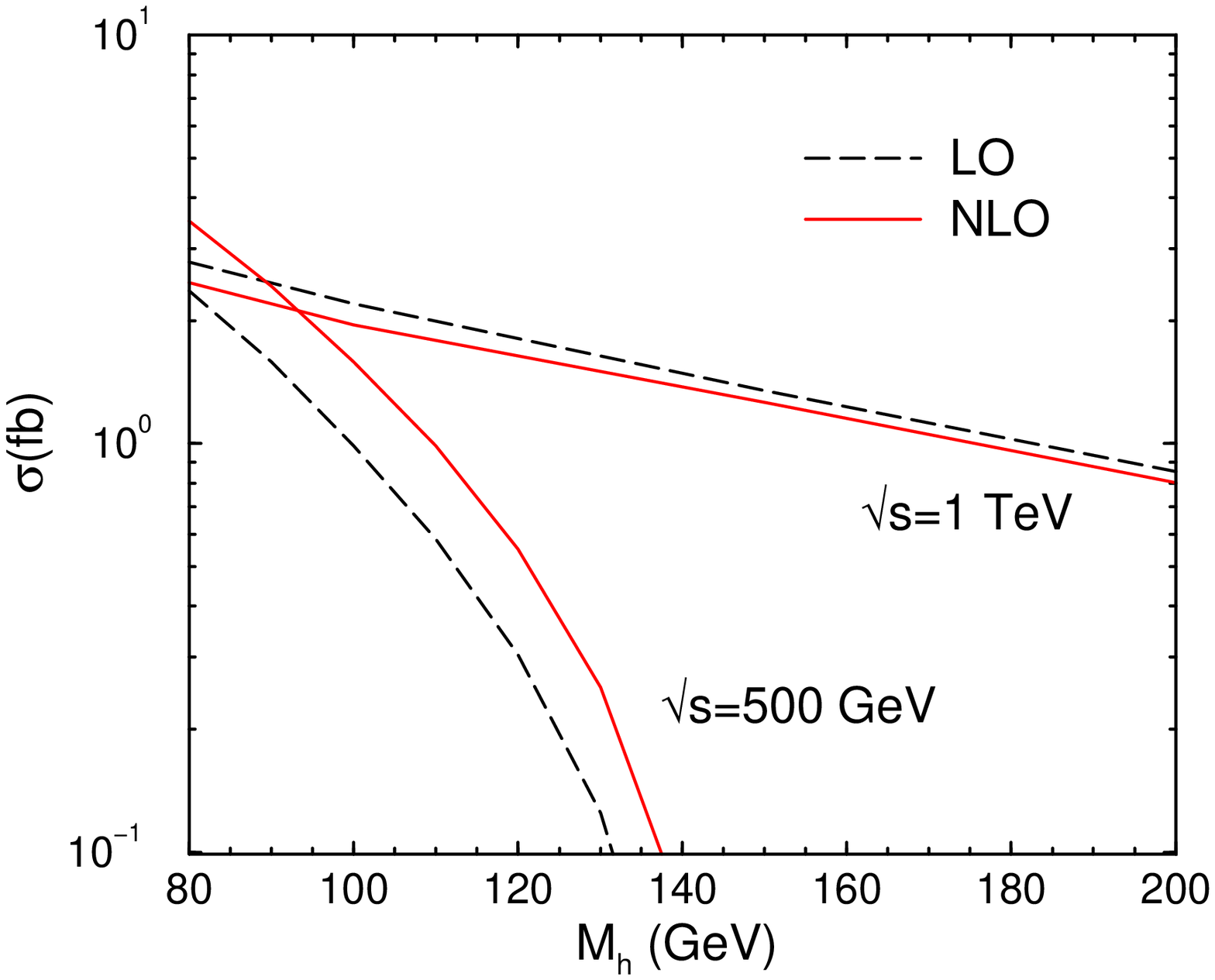}
\caption[]{\emph{Left plot}: lowest order cross section for
$e^+e^-\rightarrow t {\overline t}h$: the photon only contribution is
compared to the total cross section.  The curve labelled {\it
complete} includes both $\gamma$ and $Z$ exchange, along with
bremsstrahlung from the $Z$ boson. \emph{Right plot}: the $O(\alpha_s)$
QCD cross section is compared to the lowest order cross section.  In
both cases, curves are shown for both $\sqrt{s}=500$~GeV and
$\sqrt{s}=1~\mbox{TeV}$.  }
\label{fig:tree_and_qcd}
\end{figure}

In the Standard model ($\phi\!=\!H_{SM}$), the $t\bar tH_{SM}$ final
state can be produced both when the Higgs is radiated from a final
state top quark and when it is radiated by the intermediate $Z$
boson~\cite{sl_qcd,ditt}. In general a measurement of the inclusive
cross section would provide some mixed determination of both $g_{t\bar
tH}$ and $g_{ZZH}$ couplings. However, a more careful analysis of the
analytical properties of the cross section shows that the $ZZH_{SM}$
contribution is almost irrelevant and most of the cross sections come
from the \emph{abelian} diagrams, i.e. the diagrams where the Higgs is
radiated from the final state top quarks. Even more, they are indeed the
diagrams with a photon exchange that dominate, providing almost 98\%
of the tree level cross section ($\sigma_0$) at $\sqrt{s}\!=\!500$ GeV
and almost 90\% at $\sqrt{s}\!=\!1$ TeV~\cite{djouadi}, as illustrated
in Fig.~\ref{fig:tree_and_qcd} (left plot).

We have computed the inclusive cross section including the complete
set of $O(\alpha_s)$ QCD corrections~\cite{sl_qcd} ($\sigma_1$).
Since the photon contribution is dominant both at $\sqrt{s}\!=\!500$
GeV and at $\sqrt{s}\!=\!1$ TeV, we did not include the $Z$ diagrams
in our calculation. Reference~\cite{ditt} includes the $Z$
contributions too and gets very similar results.

We take the renormalization scale $\mu=m_t$, the top mass
$m_t\!=\!175$ GeV and the strong coupling constant
$\alpha_s(m_t^2)\!=0.11164$. The uncertainty due to the input
parameters is very small, the residual renormalization scale ($\mu$)
dependence is about 10\% and, contrary to hadronic processes, we
expect the $O(\alpha_s^2)$ QCD corrections to be small.  The results
are illustrated in Fig.~\ref{fig:tree_and_qcd} (right plot).
The effects of QCD corrections are large and positive at
$\sqrt{s}\!=\!500$ GeV, because of resonance effects, and small and
negative at $\sqrt{s}\!=\!1$ TeV. If we define the $K$-factor for this
production mode to be

\be
K(\mu) = \frac{\sigma_1}{\sigma_0} \,\,\,\,\, \mbox{then}\,\,\,\,\,
\left\{
\begin{array}{l}
K(\sqrt{s}=m_t)\simeq 1.4-2.4\,\,\,\,\, 
\mbox{for}\,\sqrt{s}=500\,\mbox{GeV}\,\,\,,\\
K(\sqrt{s}=m_t)\simeq 0.8-0.9\,\,\,\,\,
\mbox{for}\,\sqrt{s}=1\,\mbox{TeV}\,\,\,.\\
\end{array}\right.
\label{eq:kfactor}
\ee

It is interesting to note that the result for $\sqrt{s}\!=\!1$ TeV
agrees with the estimate obtained in the Effective Higgs
Approximation~\cite{sl_eha}, showing the validity of this
approximation for center of mass energies around or above
$\sqrt{s}\!=\!1$ TeV.

It is clear that the measurement of the $g_{t\bar t H}$ coupling will
be difficult at an NLC with $\sqrt{s}\!=\!500$, but it can be achieved
with higher center of mass energies.  A first study of the
background~\cite{moretti} confirmed that the $e^+e^-\rightarrow t\bar
t H_{SM}$ signal has a very distinctive signature ($W^+W^-b\bar b
b\bar b$) and offers many handles to reduce both the electroweak
($e^+e^-\rightarrow t\bar t Z\rightarrow W^+W^-b\bar bb\bar b$) and
the QCD ($e^+e^-\rightarrow t\bar t g\rightarrow W^+W^-b\bar bb\bar b
$) backgrounds.  Stimulated by discussions during this series of NLC
workshops, real simulation analyses have been recently
performed~\cite{juste,hsl_isajet}. They agree on the fact that for
center of mass energies of the order of $\sqrt{s}\!=\!1$ TeV or higher,
a precision of about 10\% can be reached on $g_{t\bar t H}$.

\subsection{Supersymmetry}
\label{subsec:ttbar_susy}
In the minimal supersymmetric model we have to consider the associated
production of a $t\bar t$ pair with both the two scalar ($h^0$, the
lighter one, and $H^0$, the heavier one) and the pseudoscalar ($A^0$)
Higgses. The processes $e^+e^-\rightarrow\bar t b H^+,~~ t \bar b H^-$
are only interesting in the small mass region where $H^\pm$ cannot
decay to $t$ nor $t$ decay to $H^\pm$ and we will not consider them
further~\cite{djouadi}.

The masses and couplings of $h^0,H^0$ and $A^0$ varies over the
minimal supersymmetry parameter space, which we decribe using the
canonical ($M_A$,$\tan\beta$) parametrization.  In the
\emph{intermediate mass region}, and varying $\tan\beta$ between 2.5
and 40, we find that the cross section for $e^+e^-\rightarrow t\bar t
h^0_i$ (for $h^0_i\!=\!h^0,H^0$) at $\sqrt{s}\!=\!500$ GeV, is almost
always larger than 0.75 fb, and grows for $\sqrt{s}\!=\!1$ TeV. On
the other hand, the cross section for $e^+e^-\rightarrow t\bar t A^0$
is always very small (e.g., less than $10^{-2}$ fb for
$\sqrt{s}\!=\!500$ GeV).

\begin{figure}[hbtp]
\centering
\epsfysize=1.5in
\leavevmode\epsffile{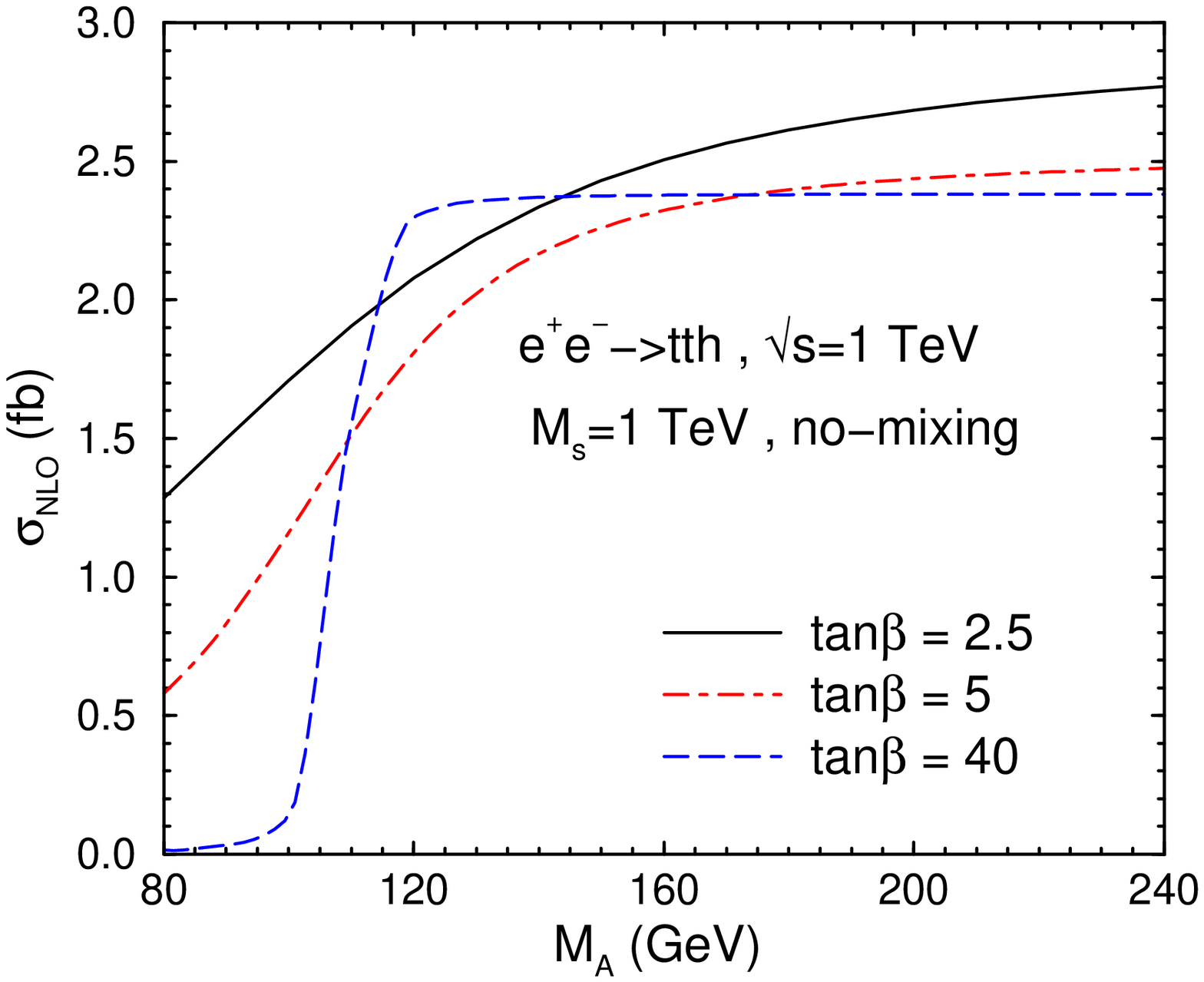}
\hspace{2.truecm}
\epsfysize=1.5in
\leavevmode\epsffile{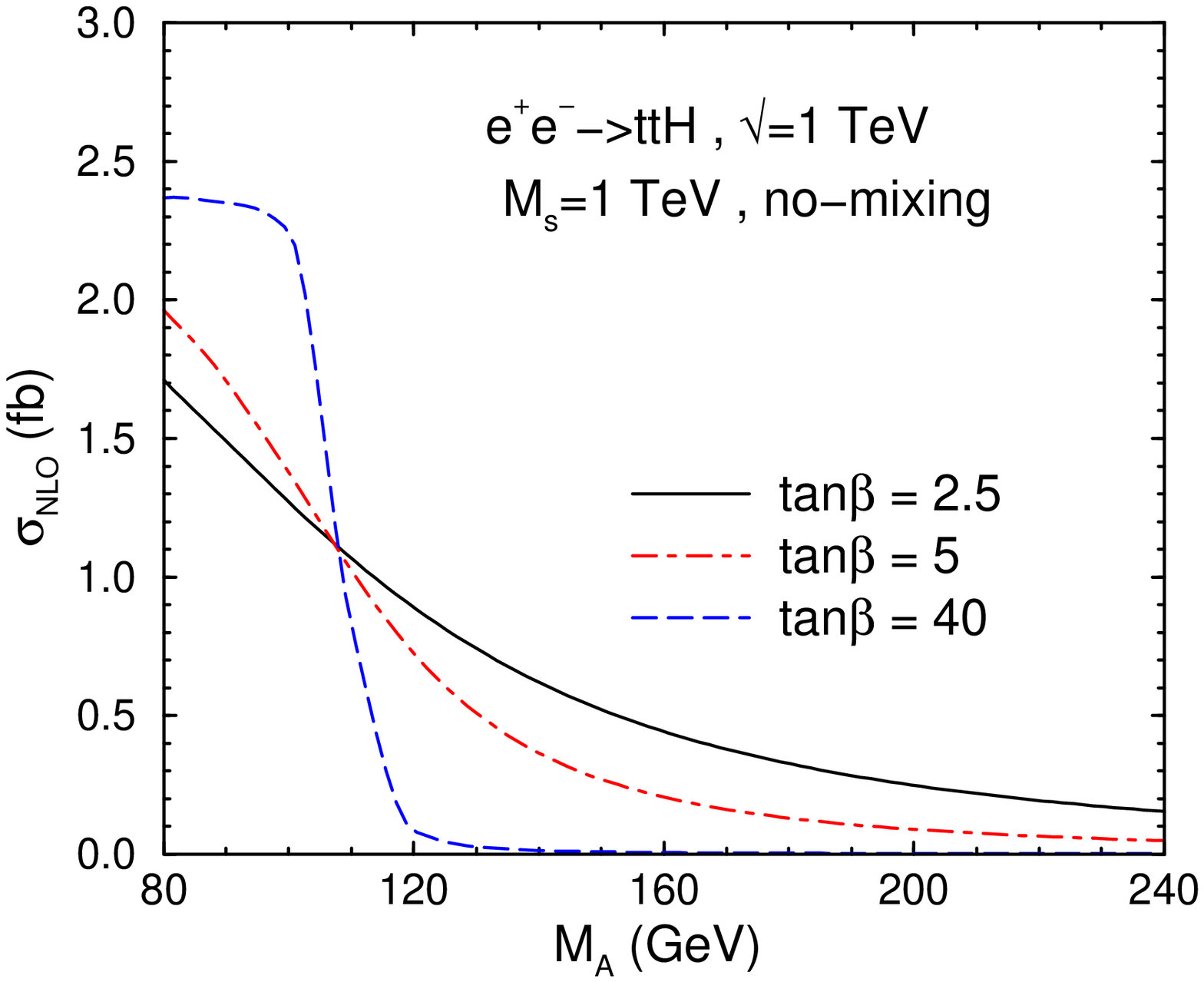}
\caption[]{Next-to-leading order result for $e^+e^-\rightarrow t
{\overline t} h^0$ (\emph{left plot}) and $e^+e^-\rightarrow t
{\overline t} H^0$ (\emph{right plot}), at $\sqrt{s}\!=\!1$~TeV using
$K\!=\!0.94$.  The squarks are taken to have a common mass,
$M_S=500~GeV$, and the scalar mixing is neglected.}
\label{fig:susy_hH_qcd}
\end{figure}

The inclusive cross section now contains a purely supersymmetric
contribution given by $e^+e^-\rightarrow Z^*\rightarrow
A^0h^0_i\rightarrow b\bar bh^0_i$, due to the $ZA^0h^0_i$
couplings~\cite{sl_susy}. This new term could spoil the sensitivity of
$t\bar th^0_i$ production to the corresponding Yukawa
couplings. Remarkably enough this does not happen for the top quark
case, and the inclusive cross section still consists almost entirely
of those diagrams in which the Higgs is radiated from the top
quark. Therefore, the precision reach estimated for the Standard Model
case still holds for the supersymmetric case. Moreover, we can compute
the $O(\alpha_s)$ cross sections by simply multiplying the tree level
supersymmetric results~\cite{djouadi,sl_susy} by the $K$-factors
calculated in the Standard Model case. As an example, the values of
$\sigma(e^+e^-\rightarrow t\bar t h^0_i)$ obtained for the
$\sqrt{s}\!=\!1$ TeV, can be read from Fig.~\ref{fig:susy_hH_qcd}.
\section{Associated production of a Higgs boson with a $b\bar b$ pair}
\label{sec:bbbar}
In the Standard Model, the $e^+e^-\rightarrow b\bar b H_{SM}$
production cross section is small and not sensitive to the $g_{b\bar b
H}$ coupling alone. Contrary to what we have seen in Section
\ref{subsec:ttbar_sm} for the $t\bar tH_{SM}$ case, the $b\bar
bH_{SM}$ production mode receives very important contributions from
$e^+e^-\rightarrow Z^*H_{SM}\rightarrow b\bar b H_{SM}$ and the
inclusive cross section measurement is therefore sensitive to a
mixture of both $g_{b\bar b H}$ and $g_{ZZH}$ couplings.

The supersymmetric scenario is however much more interesting since
having more Higgs bosons ($\phi=h^0,H^0,A^0,H^\pm$) allows more ways
to pin down the fermion-Higgs Yukawa couplings. Moreover in some
regions of the supersymmetric parameter space, as for large
$\tan\beta$, some of the $b\bar b\phi$ couplings can be very enhanced
and the $e^+e^-\rightarrow b\bar b\phi$ cross sections can be
substantially larger than in the Standard Model.
\begin{figure}[hbtp]
\centering
\epsfysize=1.5in
\leavevmode\epsffile{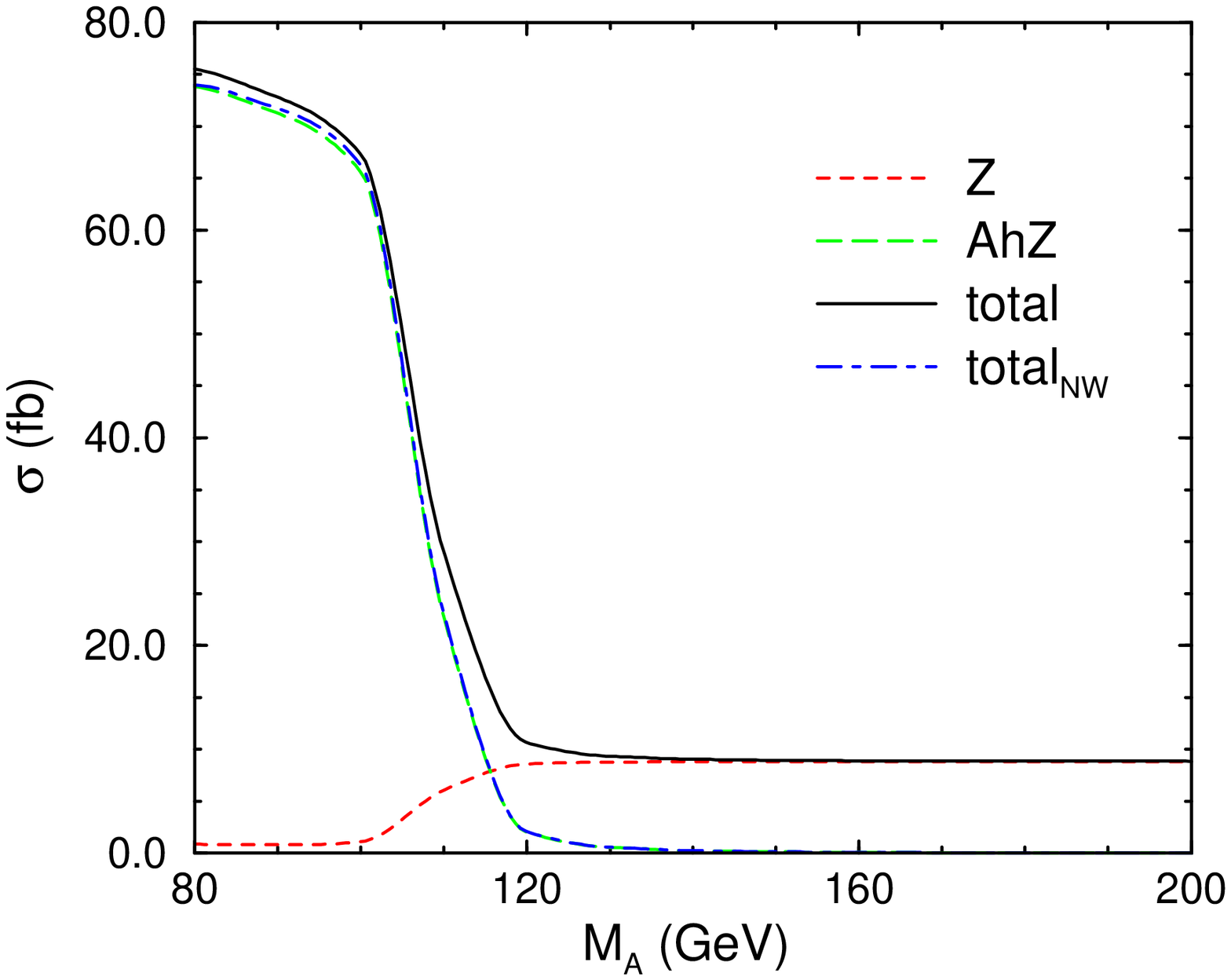}
\hspace{2.truecm}
\epsfysize=1.5in
\leavevmode\epsffile{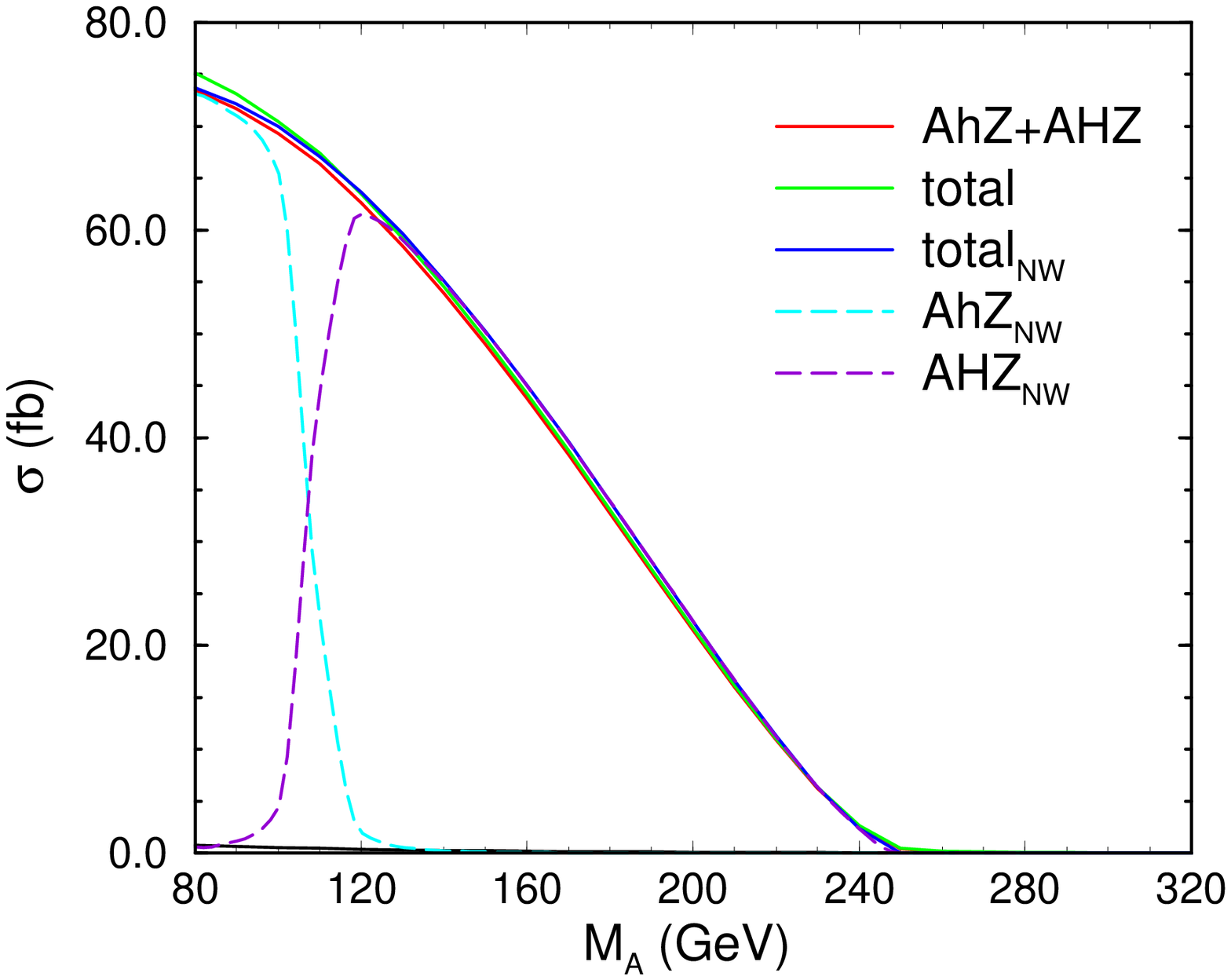}
\caption[]{Contributions to $e^+e^-\rightarrow b {\overline b}h^0$
  (\emph{left plot}) and $e^+e^-\rightarrow b {\overline b}A^0$
  (\emph{right plot}) at $\sqrt{s}\!=\!500$~GeV, for
  $\tan\beta\!=\!40$.  The curve labelled `NW' is the narrow width
  approximation of the cross section and it includes QCD corrections
  in the resonance region.  The squarks are assumed to have a common
  mass, $M_S\!=\!500$~GeV, and the scalar mixing is neglected.}
\label{fig:bbh_bbA}
\end{figure} 

Another interesting fact is that large resonant contributions affect
the genuinely supersymmetric production mode $e^+e^-\rightarrow
h^0_iA^0\rightarrow (h^0_i,A^0)b\bar b$, when $M_A\simeq M_{h_i}$,
i.e. for low values of $M_A$. In the resonant region, the $ZA^0h^0_i$
contribution is of course dominant compared to both top-Higgs and
Z-Higgs bremstrahlungs, and can increase the production rates by more
that one order of magnitude. This is true both for the scalars
$h^0,H^0$ and for the pseudoscalar $A^0$ Higgs boson, and multiple
measurements of $g_{bb\phi}g_{ZA^0h^0_i}$ can be obtained. We can work
in the narrow width approximation and include $O(\alpha_s)$ QCD
corrections by simply taking into account $O(\alpha_s)$ corrections to
the width of the resonant Higgs boson.  As soon as we move from
resonance, this is not true anymore: the Z-Higgs bremsstrahlung
contribution becomes important, even dominant, and the inclusion of
$O(\alpha_s)$ QCD corrections would require a complete
calculation. However, we are only interested in the resonant region,
since only here are the rates enhanced.

Examples of $\sigma(e^+e^-\rightarrow b\bar b h^0)$ and
$\sigma(e^+e^-\rightarrow b\bar b A^0)$ are given in
Fig.~\ref{fig:bbh_bbA}, where the $ZA^0h^0_i$ resonances, the
excellence of the narrow width approximation (NW) and some of the
previously described characteristics are evident.

\section{Conclusions}
\label{sec:concl}
The associated production of a Higgs boson with a $t\bar t$ quark pair
may allow the precision measurement of the top Yukawa coupling at a
Next Linear Collider. We have calculated the production rate at
$O(\alpha_s)$ and analyzed both the Standard Model and the minimal
supersymmetric case. In both cases a 10\% level precision is
reachable, for $\sqrt{s}\!=\!1$ TeV or higher.  The associated
production of a Higgs boson with a $b\bar b$ pair, although negligible
in the Standard Model, can be very interesting in some regions of the
minimal supersymmetry parameter space, due to enhanced couplings and
resonant contributions.

\section*{Acknowledgments}
The work of S.D. was supported by the U. S. Department of Energy under
Contract No. DE-AC02-76CH00016. The work L.R. was supported in part by
the U. S. Department of Energy under contract number DE-FG02-97ER41022.


\section*{References}
\begin{thebibliography}{99}
\bibitem{read} A.~Read, proceedings of the \emph{EPS-HEP Meeting}, Tampere,
Finland, July 1999.
\bibitem{gay} P.~Gay, proceedings of the \emph{EPS-HEP Meeting}, Tampere,
Finland, July 1999.
\bibitem{sl_qcd} S.~Dawson and L.~Reina, \Journal{\PRD}{59}{054012}{1999}.
\bibitem{sl_susy} S.~Dawson and L.~Reina,\Journal{\PRD}{60}{015003}{1999}.
\bibitem{ditt} S.~Dittmaier, M.~Kramer, Y.~Liao, M.~Spira, and P.~Zerwas, 
\Journal{\PLB}{441}{383}{1998}.
\bibitem{djouadi} A.~Djouadi, J.~Kalinowski and P.M.~Zerwas, 
\Journal{\ZPC}{54}{255}{1992}.
\bibitem{sl_eha} S.~Dawson and L.~Reina, \Journal{\PRD}{57}{5851}{1998}. 
\bibitem{moretti} S.~Moretti, \Journal{\PLB}{452}{338}{1999}.
\bibitem{juste} A.~Juste and M.~Merino, these proceedings.
\bibitem{hsl_isajet} H.~Baer, S.~Dawson and L.~Reina, hep-ph/990630,
accepted for publication in \PRD.

\end{thebibliography}
\end{document}